\documentclass[preprint,showpacs,preprintnumbers,amsmath,amssymb,nofootinbib]{revtex4-1}
\usepackage{amsmath}
\usepackage{bm}
\usepackage{times}
\usepackage{braket}
\usepackage{color}
\usepackage{graphicx}
\usepackage{epsfig}
\usepackage{slashed}
\usepackage{hyperref}
\usepackage{multirow}
\usepackage{booktabs}
\usepackage{array}
\newcommand{\beq}{\begin{eqnarray}}
\newcommand{\eeq}{\end{eqnarray}}

\def \ijmpa{ {\bf Int. J. Mod. Phys. A}  }

\def \copc{ {\bf Comput. Phys. Commun. } }
\def \epjc{{\bf Eur. Phys. J. C} }

\def \jcap{ {\bf JCAP}  }
\def \npb{ {\bf Nucl. Phys. B} }
\def \plb{ {\bf Phys. Lett. B} }

\def \prt{  {\bf Phys. Rept.} }

\def \prd{ {\bf Phys. Rev. D} }
\def \prl{ {\bf Phys. Rev. Lett.}  }

\def \jhep{ {\bf JHEP}  }

\definecolor{Red}{rgb}{1.,0.,0.}

\definecolor{Blue}{rgb}{0.,0.,1.}

\definecolor{nicered}{rgb}{0.7,0.1,0.1}
\definecolor{nicegreen}{rgb}{0.1,0.5,0.1}
\def\lsim{ {\ \lower-1.2pt\vbox{\hbox{\rlap{$<$}\lower6pt\vbox{\hbox{$\sim$}}}}\ } }
\def\gsim{ {\ \lower-1.2pt\vbox{\hbox{\rlap{$>$}\lower6pt\vbox{\hbox{$\sim$}}}}\ } }

\hypersetup{colorlinks,citecolor=nicegreen,linkcolor=nicered}
\begin{document}
\title{Search for single production of a top quark partner \\[0.15cm] via the $T\to th$ and $h\to WW^{\ast}$ channels at the LHC}
\author{Yao-Bei Liu\footnote{E-mail: liuyaobei@hist.edu.cn}}
\affiliation{Henan Institute of Science and Technology, Xinxiang 453003, P.R.China }
\affiliation{School of Physics \& Astronomy, University of Southampton, Highfield, Southampton SO17 1BJ, UK}
\author{Stefano Moretti\footnote{E-mail: s.moretti@soton.ac.uk}}
\affiliation{School of Physics \& Astronomy, University of Southampton, Highfield, Southampton SO17 1BJ, UK}
\affiliation{Particle Physics Department, Rutherford Appleton Laboratory, Chilton, Didcot, Oxon OX11 0QX, UK}

\vspace*{1truecm}
\begin{abstract}
Many scenarios of  physics beyond the Standard Model  that address the hierarchy problem also predict the existence of vectorlike top quark partners, which are generally expected around the TeV scale. In this paper, we propose to search for a vectorlike top quark partner with charge $2/3$ in a simplified model including only two free parameters, the coupling constant  $g^{\ast}$ and  top quark partner mass $m_T$. We investigate the observability of the top quark partner through the process $pp \to T(\to th)j\to t(\to b W^+\to b \ell^{+} \nu_{\ell})h( \to WW^{\ast}\to \ell^{+}\nu\ell^{-}\bar{\nu}) j$,
where $T$ is the heavy top quark partner and $h$ the SM-like Higgs boson, at the  Large Hadron Collider (LHC). The discovery prospects and exclusion limits on the parameter
plane defined by ($m_T,g^{\ast}$) are obtained for the already scheduled LHC runs
as well as at the future High-Luminosity~LHC (HL-LHC). The constraints and projected sensitivities are also interpreted in a realistic model, i.e., the minimal Composite Higgs Model with singlet top quark partners.
Finally, we also analyze the projected sensitivity in terms of the production cross section times branching fraction at the (HL-)LHC.
\end{abstract}

\maketitle
\newpage

\section{Introduction}
 To solve the gauge hierarchy problem, many extensions of the Standard Model (SM) predict top quark partners, which play an important role in canceling potentially large top quark loop corrections to the Higgs boson mass~(for a review, see~\cite{hmass}). Vectorlike top quark partners $T$'s with the same color and electroweak (EW) quantum numbers as the top quark ones have been introduced in many new physics (NP) scenarios, such as little Higgs models~\cite{littlehiggs}, extra dimensions~\cite{ED}, twin Higgs models \cite{twin-higgs}, and composite Higgs models (CHMs)~\cite{Agashe:2004rs}. In general, these new particles are at or just below the TeV scale and might generate characteristic signatures at  current and future high energy colliders. In particular, the discovery of these  top quark partners at the Large Hadron Collider (LHC) will be very important to test these NP models.

At the LHC,  vectorlike top quark partners can be produced in pairs or singly, both of which have been widely studied via various final states in the literature: see, e.g.,  \cite{p1,p2,p3,p4,p5,p6,p7,p8,p9,p10,p11,jhep1304-004,prd-88-094010}. While for light $T$ states their pair production is vastly dominant, for heavy top quark partners, the single channel  mode eventually dominates over pair production due to  a larger phase space. Vectorlike top quark partners generally only mix with the third generation SM quarks~\cite{prd-90-014007}, but, in some models, they can mix with the light SM quarks generations too, which opens up new production mechanisms and makes the investigation of such new particles at the
LHC very promising~\cite{bound-tp,1007.2933,Buchkremer:2013bha,jhep-1108-080,jhep-1502-032,lyb-2017,tp-exotic-decay}.
Given the current constraints from direct searches by the ATLAS and CMS Collaborations with an integrated luminosity of 35--36 fb$^{-1}$, the minimum mass of a top quark partner is set at about 1.2--1.3 TeV, for a variety of signatures via the pair production processes~\cite{13tev-tp1,13tev-tp2}.
Very recently, the ATLAS Collaboration presented a search optimized for a singly produced
 vectorlike $T$ quark at $\sqrt{s}=13$ TeV via the $T\to bW$ channel with the $W$ boson decaying leptonically~\cite{1812.07343}. The results show that, for the $T$ quark mass range of 800~GeV to 1200~GeV, the upper exclusion limit on the $TWb$ coupling strength $C_{L}^{Wb}$ is $0.25-0.49$.

The High-Luminosity LHC (HL-LHC) is expected to reach 3000 fb $^{-1}$~\cite{HL-LHC}, which will be very beneficial for discovering possible new physical signals even for small production and/or decay rates. Hence, at such a high luminosity, a variety of $T$ decay channels can, in principle, be accessed.
In the past few years, the discovery of a SM-like Higgs boson $h$~\cite{atlas-cms-higgs} has rendered the $T\to th$ decay channel promising, so it
has been considered as a $T$ search mode, wherein the SM-like Higgs boson    decays to $h\to b\bar{b}$~\cite{tp-th-bb1,tp-th-bb2,tp-th-bb3}, $h\to \gamma\gamma$~\cite{tp-th-aa}, and $h\to ZZ$~\cite{tp-th-zz}. As we know, the $h\to WW^{\ast}$ decay channel has the second largest branching ratio (BR), of about $22\%$, and also has the advantage of  a smaller backgrounds than
$h\to b\bar{b}$ (which is indeed the dominant mode). This encourages us to further analyze the $T\to th$ decay channel followed by the pure leptonic mode $h\to WW^{\ast}\to \ell^{+}\nu\ell^{-}\bar{\nu}$ in order to eventually provide a  sensitivity comparable to that of other modes for the  (HL-)LHC. Assuming single-$T$ production, for the hadronic and leptonic decay of the top quark, there are two cases for the final state, namely, two leptons plus multijets and trilepton signals, but the former will suffer from the large SM background coming from the $t\bar{t}+ \text{jets}$ process.
Therefore, we  study here the observability of  single-$T$
 production at the (HL-)LHC via the $T\to t(\to b W^+\to b\ell^+\nu_\ell) h (\to {WW^{\ast}}\to \ell^{+}\nu\ell^{-}\bar{\nu})$ decay channel, accompanied by at least one jet, $j$.
 (It should be noted that our results are model independent and can be applied to several NP scenarios, including those with  singlet top quark partners.)

This paper is organized as follows. In Sec.~II we systematically analyze the signals and backgrounds for the single top quark partner production process in a simplified model, which only comprises two independent parameters, as well as present our strategy to determine the reconstructed masses for the Higgs boson and top quark partner, including discussing the exclusion and discovery potential at the  (HL-)LHC. Finally, we present our conclusions in Sec.~III.

\section{Searches for Top Partners at the HL-LHC}
\subsection{A simplified model including a singlet top quark partner}
As proposed in
Refs.~\cite{1007.2933,Buchkremer:2013bha},  vectorlike top quark partners could be embedded in different representations of the weak $SU(2)$ group. Here we consider an $SU(2)$ singlet vectorlike $T$ quark with charge 2/3. In many cases, such vectorlike top quark partners share similar final state
topologies with different BRs and single production couplings.
Thus, it is favourable to use simplified model approaches in searching for the possible signals of top quark partners at the LHC, which only include the mass of the top quark partner and its single production coupling as free
parameters.
A generic parametrization of an effective Lagrangian for top quark partners is given by~(for some details one can see Refs.~\cite{1007.2933,Buchkremer:2013bha})
\beq
{\cal L}_{\rm eff} =&& \frac{gg^{\ast}}{2\sqrt{2}}[\bar{T}_{L}W_{\mu}^{+}
    \gamma^{\mu} b_{L}+
    \frac{g}{\sqrt{2}c_W}\bar{T}_{L} Z_{\mu} \gamma^{\mu} t_{L}
    - \frac{m_{T}}{\sqrt{2}m_{W}}\bar{T}_{R}ht_{L} -\frac{m_{t}}{\sqrt{2}m_{W}} \bar{T}_{L}ht_{R} ]+ h.c.,
  \label{TsingletVL}
\eeq
where $g$ is the SM $SU(2)$ gauge coupling
  constant, $c_W=\cos\theta_{W}$, and $\theta_W$ is the Weinberg angle.
\begin{figure}[ht]
\begin{center}
\vspace{0.5cm}
\centerline{\epsfxsize=9cm \epsffile{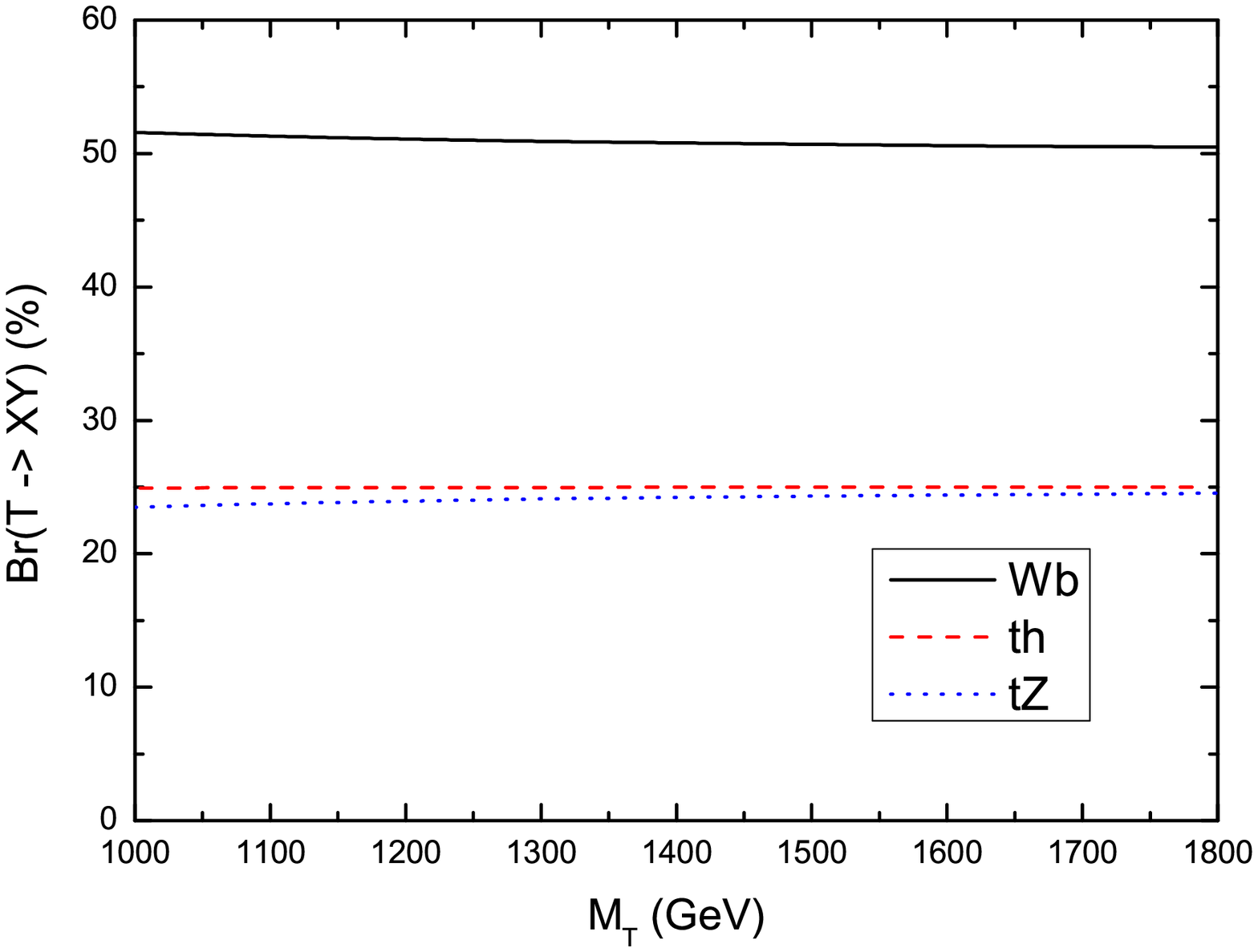}\hspace{-1.0cm}\epsfxsize=9cm \epsffile{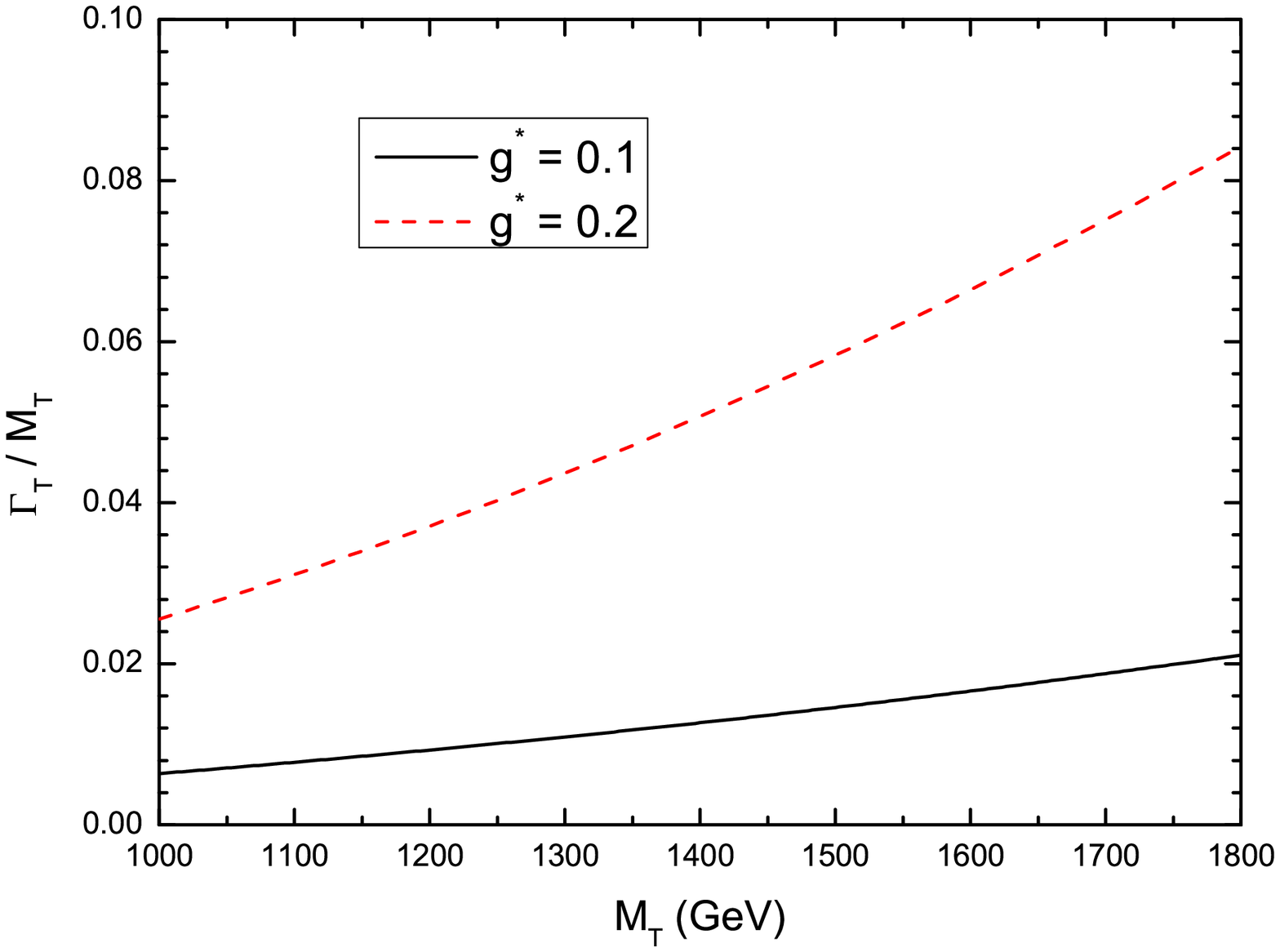}}
\caption{BRs (left) and decay widths (right) of the top quark partner as a function of its  mass.}
\label{br}
\end{center}
\end{figure}

From Eq.~(\ref{TsingletVL}), one can see that there are indeed only two free parameters, the top quark partner mass $m_T$ and the coupling strength to SM quarks in units of the SM coupling $g$, $g^{\ast}$. The tree level decay widths of the top partner into SM objects and their large mass limits are given in the Appendix~\ref{sec:decay}.
In Fig.~\ref{br}, we show the BRs of three decay channels $T\to$  $bW$, $tZ$, and $th$ as well as their  decay widths by varying the top quark partner mass at fixed $g^{\ast}$.
 One can see that BR$(T\to th)\approx {\rm BR}(T\to tZ)\approx \frac{1}{2}{\rm BR}(T\to Wb)$
 is a good approximation as expected from the Goldstone
boson equivalence theorem~\cite{ET-hjh,ET-hjh1}. Further, the width of the top quark partner is very small with respect to its mass. Thus, it is possible to factorize the production and decay parts of the scattering amplitudes
and write the cross section as
$\sigma_T\times {\rm BR}(T\to XY)$ for a generic channel, where $\sigma_T$ is the single-$T$ production cross section and {\rm BR}$(T\to XY)$ the decay rate into the generic $XY$ final state.

\subsection{Event generation and cut flow}
\begin{figure}[htb]\vspace{-1.5cm}
\begin{center}
\centerline{\hspace{7.5cm}\epsfxsize=20cm \epsffile{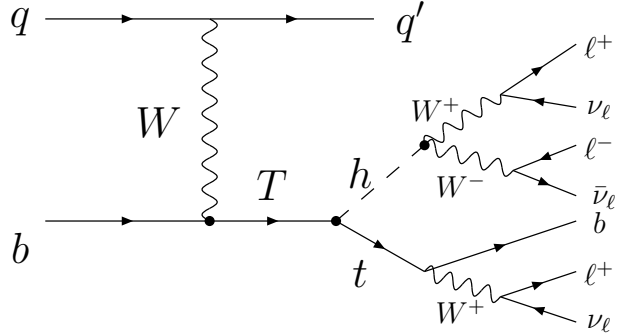}}
\vspace{-21cm}
\caption{The Feynman diagram for the production of a single-$T$ quark (and a jet) including the decay chain $T\to t(\to b\ell^{+} \nu)h(\to WW^{\ast}\to \ell^{+}\nu\ell^{-}\bar{\nu})$. }
\label{fey}
\end{center}
\end{figure}

In this subsection, we analyze the LHC observation potential by performing a
Monte Carlo (MC) simulation of the signal plus background events and explore
 the sensitivity to the top quark partner at the (HL-)LHC through the process
 \beq
 pp\to T(\to th)j \to t(\to bW\to b\ell^{+} \nu_{\ell})h(\to WW^{\ast}\to \ell^{+}\nu\ell^{-}\bar{\nu}) j.
\eeq
The Feynman diagram of the
production and decay chain is presented in Fig.~\ref{fey}.
The QCD next-to-leading order~(NLO)
production cross section of the process $pp\to Tj$ is calculated in
Ref.~\cite{NLO-tp}. From there we take a $K$-factor of 1.2 for the signal before  event generation (i.e., inclusively).
In the remainder of the paper, we will adopt three benchmark values for the $T$ mass, i.e., $m_T=$ 1.0, 1.2 and 1.5 TeV (which we will refer to in the legends as $T_{1000}$, $T_{1200}$ and $T_{1500}$, respectively).

All signal and background events are simulated at the LO by using
MadGraph5-aMC$@$NLO \cite{mg5} with the NN23LO1 parton distribution function (PDF) set~\cite{cteq}, with default renormalization and factorization
scales. The parton shower and the fast detector simulations are done with PYTHIA 8~\cite{pythia8} and DELPHES 3~\cite{delphes}, respectively.
Finally, event reconstruction is performed with MadAnalysis5~\cite{ma5}, where the anti-$k_{t}$ algorithm~\cite{antikt} is used with a radius parameter $R=0.4$ in order to select jets. Finally, we use $\sqrt s=14$ TeV in all our plots as LHC energy.

For the leptonic decay of the top quark and the full leptonic Higgs decay mode, the typical signal is three charged leptons $\ell(=e,\mu)$, one $b$-jet, one forward jet and missing transverse energy, $\slashed E_T$.
The backgrounds that can give three leptons in the final states that are considered in this analysis are: $t\bar{t}V$ ($V=W, Z$), $t\bar{t}h$ and $WZjj$. The $t\bar{t}+\text{jets}$  process, which has large cross section, may also contribute to the  background if the  third lepton comes from a $B$-hadron semileptonic decay inside a $b$-jet.
 We do not consider other backgrounds from $t\bar{t}t\bar{t}$, triboson events and $thj$, though, because their cross sections are negligible after applying  our selection cuts (see below).
  Further, we do not consider jets faking electrons either because the corresponding rates are negligible in multilepton analyses (at the level of $10^{-4}$ after selection cuts)~\cite{jetfake}.
Like for the signals, the cross sections of these backgrounds at LO are adjusted to NLO by means of
$K$-factors, which are about 1.3 for $t\bar{t}V~(V=W^{\pm},Z)$~\cite{nlo-ttv}, 1.24 for $t\bar{t}h$~\cite{nlo-tth} and 0.86 for $WZjj$~\cite{nlo-wzjj}.  The
dominant top pair production cross section is normalized to the next-to-NLO (NNLO) (in QCD) \cite{1303.6254}.

In our MC simulation, the following acceptance cuts are enforced for all signal and background events.
 \begin{itemize}
\item
Basic cuts: $p_{T}(\ell) > 10 \rm ~GeV$, $p_{T}(j, b) > 15 \rm ~GeV$, $|\eta_{\ell,b}|<2.5$, $|\eta_{j}|<5$,  $\Delta R_{bj,b\ell,\ell j}> 0.4$.
\end{itemize}

\begin{figure}[htbp]
\begin{center}
\centerline{\hspace{2.0cm}\epsfxsize=11cm\epsffile{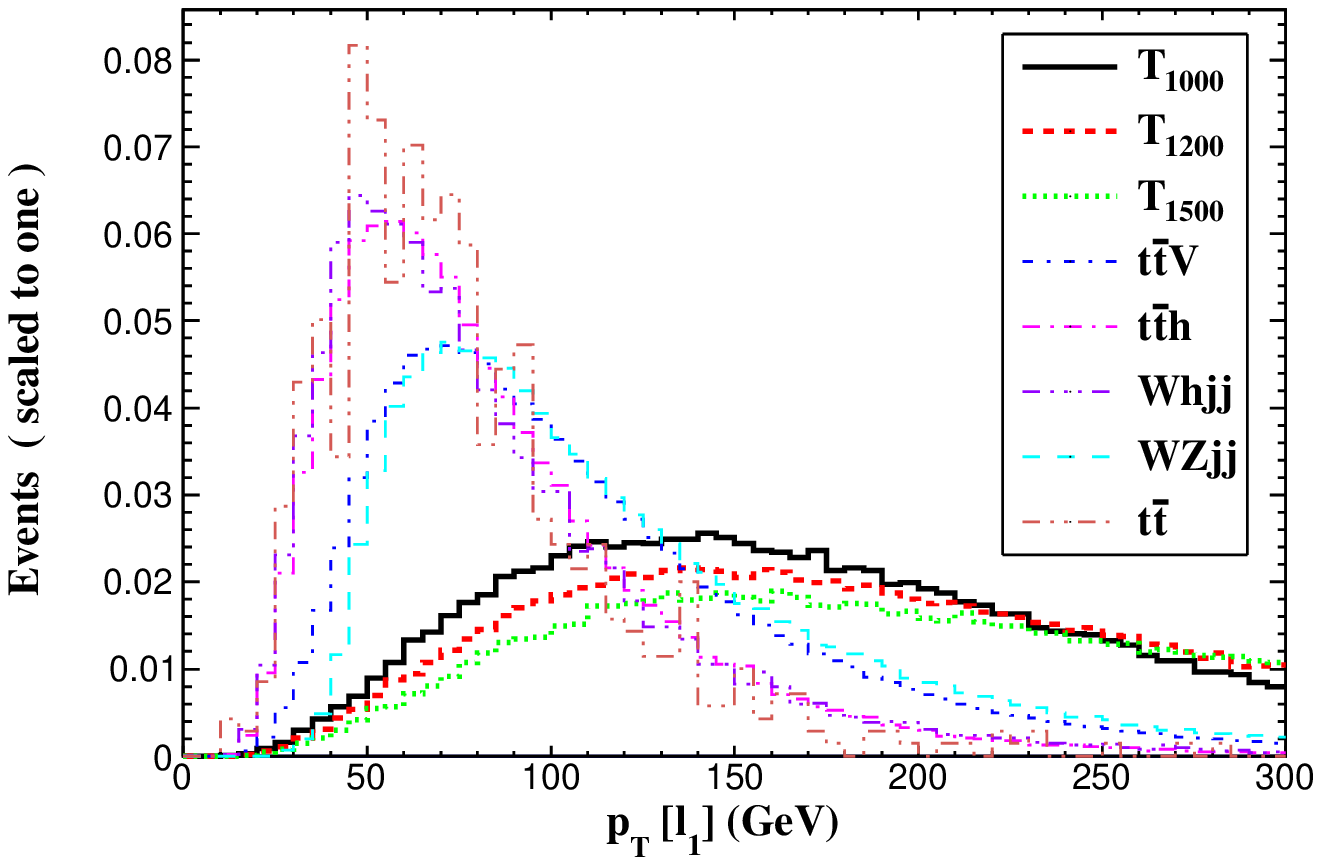}
\hspace{-3.0cm}\epsfxsize=11cm\epsffile{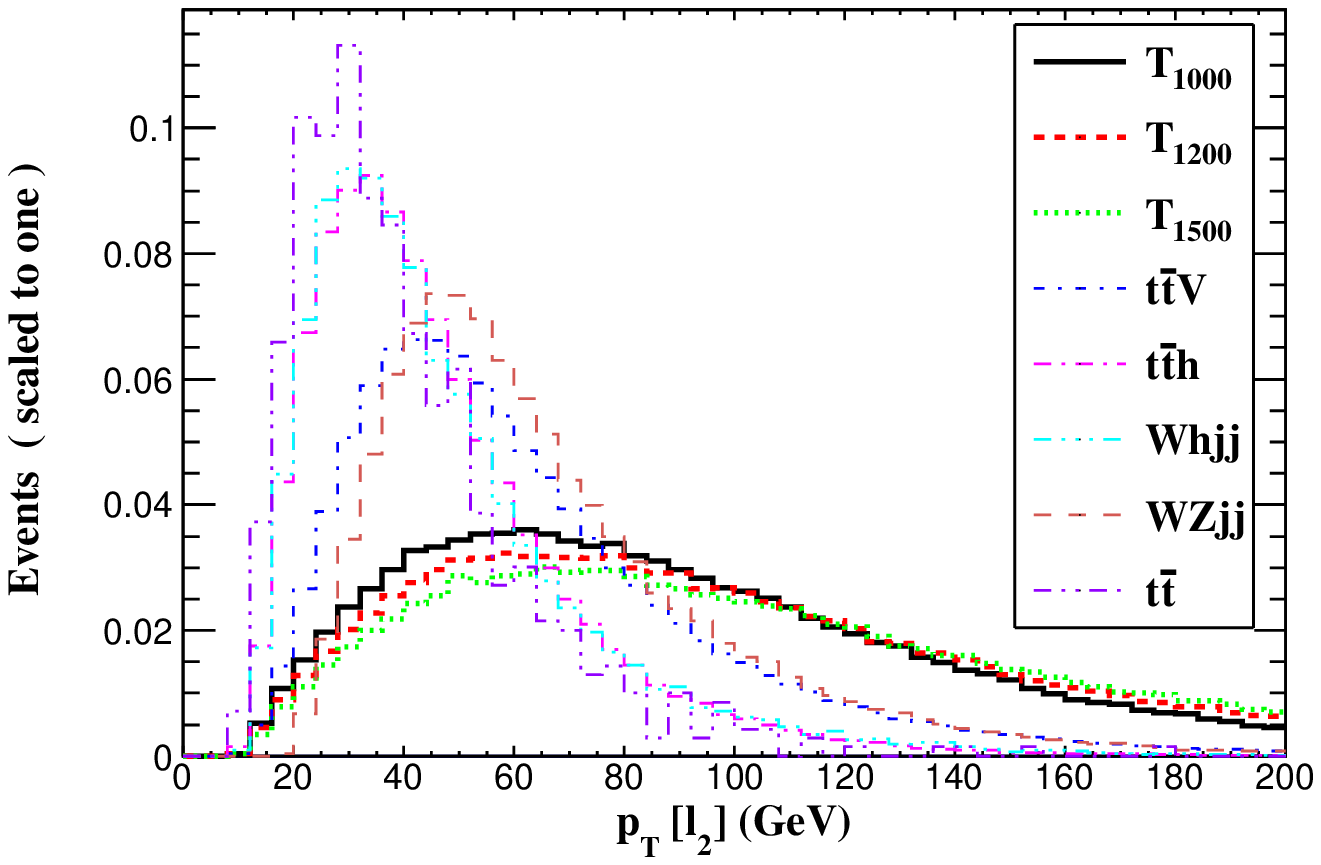}}
\centerline{\hspace{2.0cm}\epsfxsize=11cm\epsffile{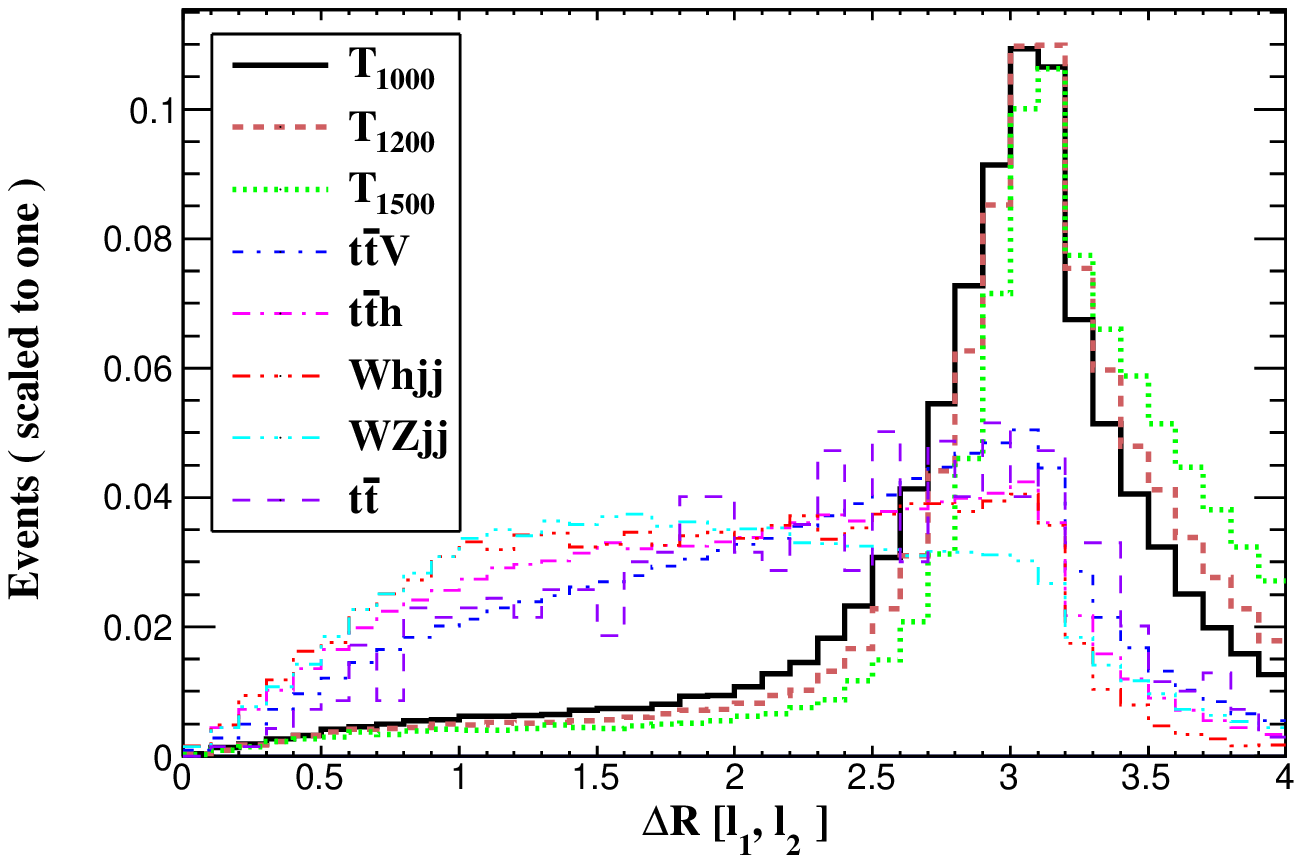}
\hspace{-3.0cm}\epsfxsize=11cm\epsffile{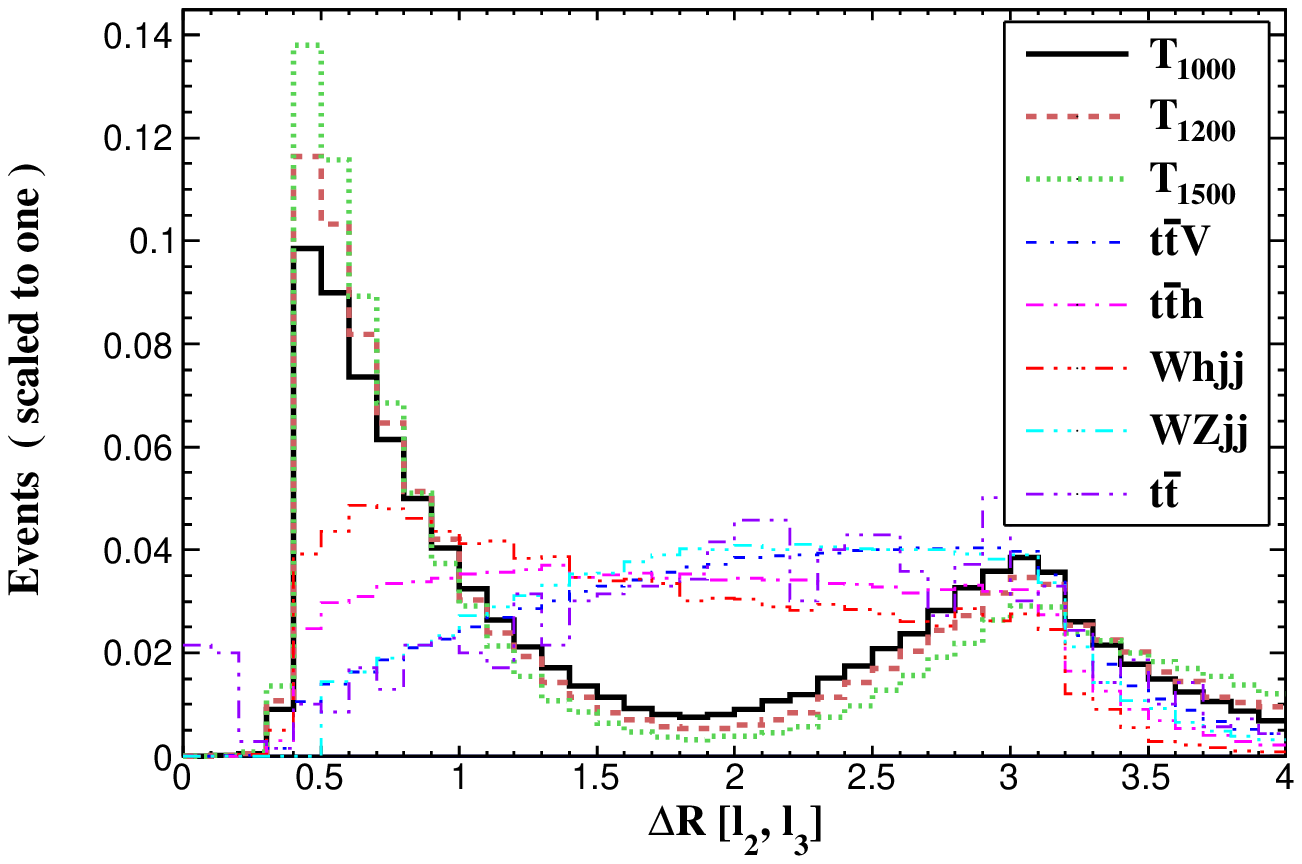}}
\caption{Normalized distributions in transverse momentum and cone separation for the signals and backgrounds.}
\label{cut-1}
\end{center}
\end{figure}

In order to choose appropriate selection cuts, in Fig.~\ref{cut-1}, we show some key normalized distributions for the signals and  backgrounds, such as (some of) the transverse momenta
$p_{T}(\ell_{i})$ and cone separations  $\Delta R(\ell_i,\ell_j)$, for all $i\neq j=1,2,3$ leptons ordered in decreasing energy. Based on these kinematical distributions, we impose the following selection cuts.
 \begin{itemize}
\item
Cut 1: Exactly three isolated leptons~[$N(\ell)=3$], with $p_{T}(\ell_{1}) > 100 \rm ~GeV$ and $p_{T}(\ell_{2}) > 25 \rm ~GeV$, and at least two jets, one of which is an isolated $b$-jet~[$N(b)= 1$]. Since the (most energetic) first lepton $\ell_{1}$ is assumed to originate from the leptonically decaying top quark, we require $\Delta R(\ell_{1}, \ell_{2})> 2.5$ and $\Delta R(\ell_{2},\ell_{3})< 1$.
\end{itemize}

\begin{figure}[!htb]
\begin{center}
\vspace{0.5cm}
\centerline{\hspace{1.5cm}\epsfxsize=13cm \epsffile{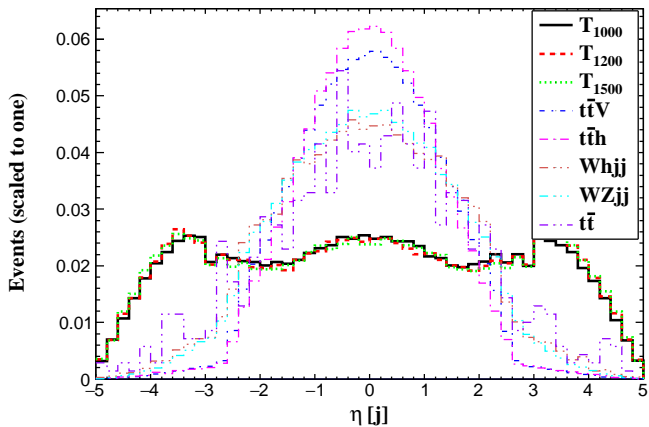}}
\caption{Normalized distribution in  pseudorapidity of the forward/backward jet for the signals and backgrounds.}
\label{etaj}
\end{center}
\end{figure}

The extra jet (from a valence quark emission) entering the signal final state
always has a strong forward/backward nature, which is a useful handle in suppressing the SM backgrounds. The distribution of the pseudorapidity of the forward/backward jet is plotted in Fig.~\ref{etaj}
for the signals and backgrounds. Based on this spectrum, one can further reduce the
 backgrounds through the following cut.
\begin{itemize}
\item Cut 2: The light untagged jet is required to have $\mid\eta_{j}\mid > 2.4$.
\end{itemize}

\begin{figure}[!htb]
\begin{center}
\centerline{\hspace{2.5cm}\epsfxsize=11cm\epsffile{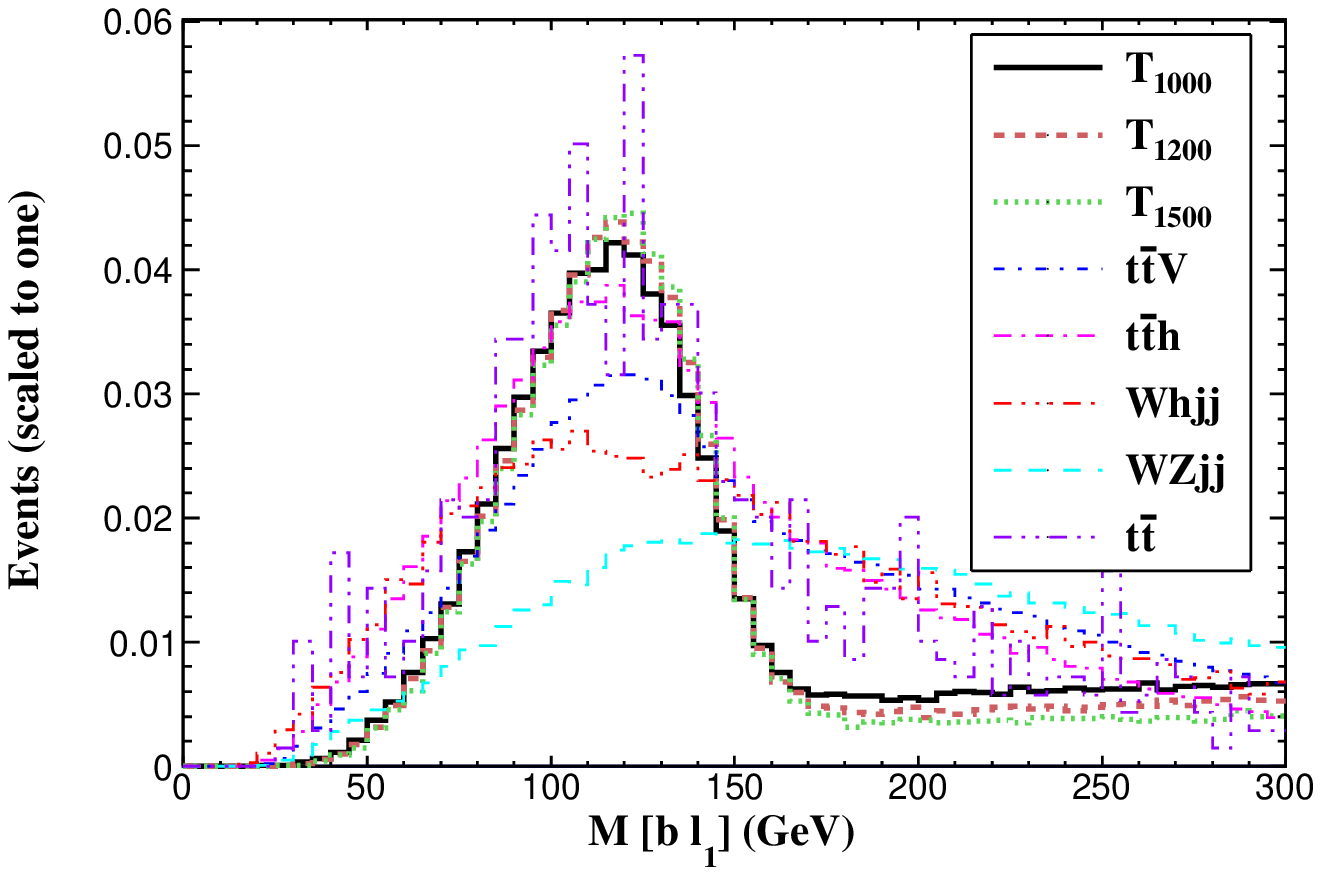}
\hspace{-3.0cm}\epsfxsize=11cm\epsffile{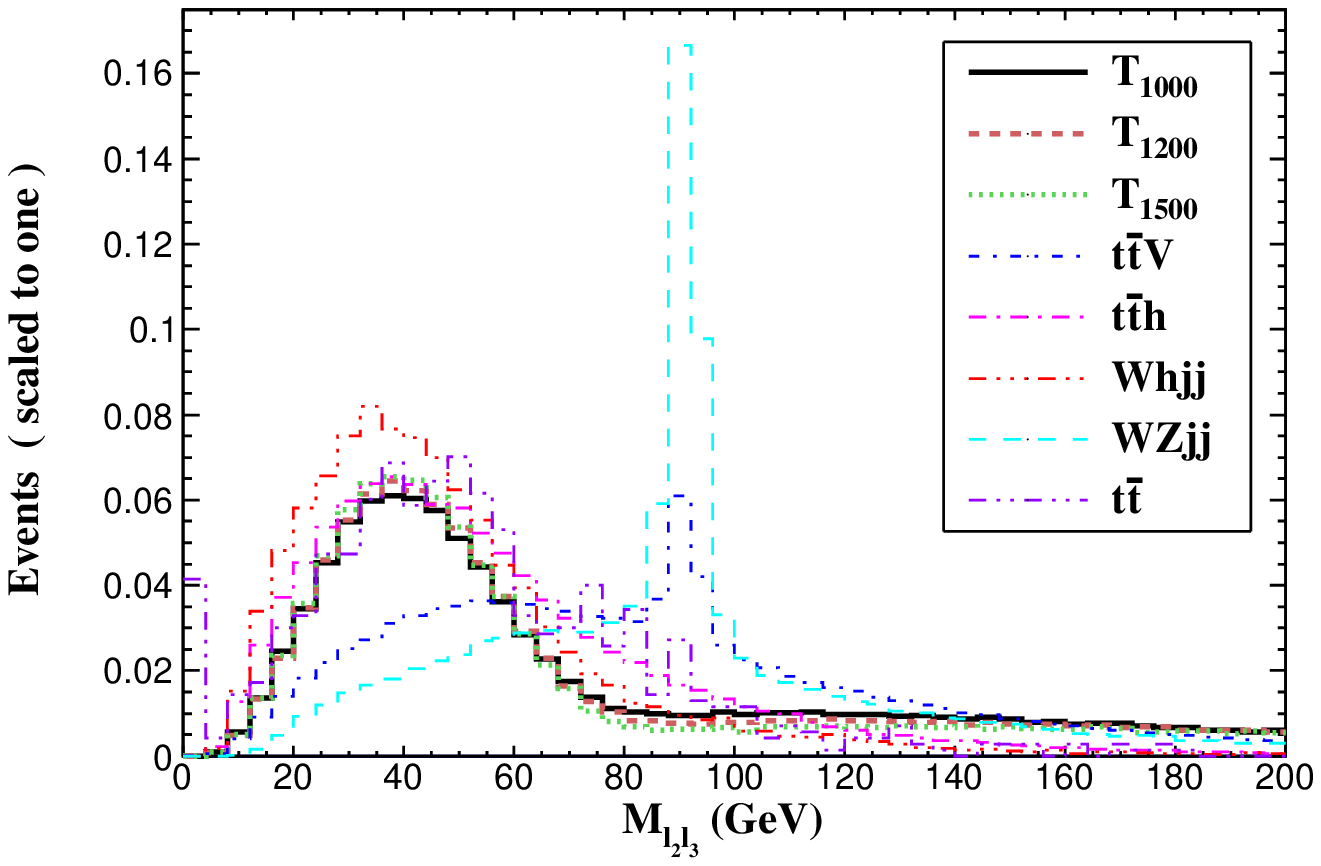}}
\caption{Normalized  distributions in  invariant mass
 of the $b\ell_{1}$ and $\ell_{2}\ell_{3}$ systems for the signals and backgrounds.}
\label{cut-mtmh}
\end{center}
\end{figure}

The invariant mass of the $b\ell_{1}$ and $\ell_{2}\ell_{3}$ systems is plotted in Fig.~\ref{cut-mtmh} for the  signals
 and  backgrounds.  One can see that, for $T$ events,  the invariant mass of the $b$-jet and the leading lepton $M_{b\ell_{1}}$ is always less than the top quark mass since the tagged $b$-jet and leading lepton in our signals come from the same top quark decay. A similar feature also appears for the invariant mass of the $\ell_{2}\ell_{3}$ system, which is very different from the resonant $Z$ boson one typical of most SM noise. Thus we can further reduce the backgrounds via the following cuts.
\begin{itemize}
\item
Cut 3: $M_{b\ell_{1}}< 150 \rm ~GeV$.
\item
Cut 4: $13 {\rm ~GeV} < M_{\ell_{2}\ell_{3}}< 60 \rm ~GeV$.
\end{itemize}

\begin{figure}[htb]
\begin{center}
\vspace{-0.5cm}
\centerline{\hspace{2.5cm}\epsfxsize=14cm \epsffile{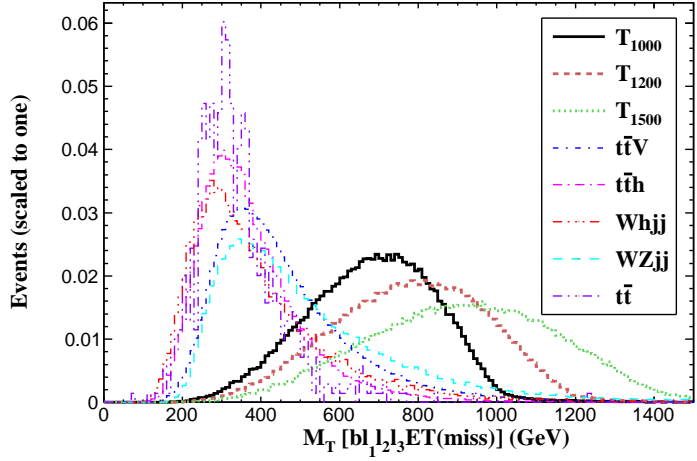}}
\caption{Normalized distribution in cluster transverse mass
of  the $b \ell_{1}\ell_{2}\ell_{3} \slashed E_T$ system for the signals and backgrounds.}
\label{mtp}
\end{center}
\end{figure}

To reconstruct the top quark partner mass, we use a cluster transverse mass, defined as~\cite{epjc-74-3103}
\beq
M_{T}^{2}(b\ell_{1}\ell_{2}\ell_{3}\slashed E_T)=(\sqrt{p_{T}^{2}(b\ell_{1}\ell_{2}\ell_{3})+M^{2}_{b\ell_{1}\ell_{2}\ell_{3}}}+\slashed E_T)^{2}-(\vec{p}_{T}(b\ell_{1}\ell_{2}\ell_{3})+\slashed E_T)^{2},
\eeq
where $\vec{p}_{T}(b\ell_{1}\ell_{2}\ell_{3})$ is the total transverse momentum of all  visible particles (but the forward/backward jet) and $M_{b\ell_{1}\ell_{2}\ell_{3}}$ is their invariant mass. In Fig.~\ref{mtp}, we show the transverse mass distribution $M_T(b\ell_{1}\ell_{2}\ell_{3}\slashed E_T)$. From this figure, we can see that the transverse mass distribution has an end point around the top quark partner mass in the signal, unlike the backgrounds, which can then be used in the following cut to further remove SM  noise.
\begin{itemize}
\item Cut 5: $M_{T}(b\ell_{1}\ell_{2}\ell_{3}\slashed E_T)>600 \rm~GeV$.
\end{itemize}

\begin{table}[htb]
\centering %
\caption{The cut flow of the cross sections (in $10^{-3}$ fb) for our signals and the relevant
backgrounds at the LHC with $\sqrt s=14$ TeV. Here we take the gauge parameter as $g^{\ast}=0.2$. \label{cutflow}}
\vspace{0.2cm}
\begin{tabular}{|c|c|c|c|c|c|c|c|c|}
\toprule[1.0pt]
\multirow{2}{*}{Cuts}& \multicolumn{3}{c| }{Signals}&\multicolumn{5}{c|}{Backgrounds} \\ \cline{2-9}
&{1.0~TeV} & {1.2~TeV} & {1.5~TeV} & $t\bar{t}+X$ & $t\bar{t}V$ & $t\bar{t}h$ & $WZjj$ &$Whjj$\\  \cline{1-9}\hline
\hline
Basic cuts& 24&11&3.5&$1.6\times 10^{7}$&8400&240&$5.1\times 10^{4}$ &98 \\ \hline
Cut 1&3.3&1.5&0.5&24&16.2&0.3&10&0.08\\ \hline
Cut 2&1.9&0.84&0.28&4.1&0.55&0.01&10&0.007\\ \hline
Cut 3&1.7&0.73&0.25&1.1&0.36&0.009&0.15&0.007\\ \hline
Cut 4&1.4&0.6&0.21&0.51&0.15&0.005&0.024&0.002\\ \hline
Cut 5&1.3&0.58&0.2&0.05&0.018&$7.4\times 10^{-4}$&$9.7\times 10^{-4}$&$1.1\times 10^{-4}$\\ \hline
\bottomrule[1.0pt]
\end{tabular}
\end{table}

In Table~\ref{cutflow}, we show the cut flow of the signal and background cross sections after each selection  for $g^{\ast}=0.2$ and our three benchmark top quark partner masses.
One can see that the backgrounds are suppressed very efficiently after imposing all listed cuts.

\subsection{Analysis and results}

As there are only a few events for both  signals and
backgrounds after the kinematics cuts, assuming any (HL-)LHC luminosity,  we estimate the discovery ($D$) prospects and exclusion ($E$) limits using the formulas~\cite{ss-epjc}
\beq
\label{eq:ss5}
\mathcal{Z}_{D}&=&\sqrt{2\times\mathcal{L}_{\rm int}\big [(\sigma_S+\sigma_B)\ln(1+\frac{\sigma_S}{\sigma_B})-\sigma_S \big ]},\\
\label{eq:ss2}
\mathcal{Z}_{E}&=&\sqrt{-2\times\mathcal{L}_{\rm int}\big [\sigma_B\ln(1+\frac{\sigma_S}{\sigma_B})-\sigma_S \big ]},
\eeq
where $\sigma_S$ and $\sigma_B$ are the cross sections of each signal ($S$) and total background ($B$) after all cuts
 and $\mathcal{L}_{\rm int}$ is the integrated luminosity. Clearly, the values of $\mathcal{Z}_{D,E}$ are dependent on the coupling parameter $g^{\ast}$ and the top quark partner mass.

\begin{figure}[htb]
\begin{center}
\centerline{\epsfxsize=11cm \epsffile{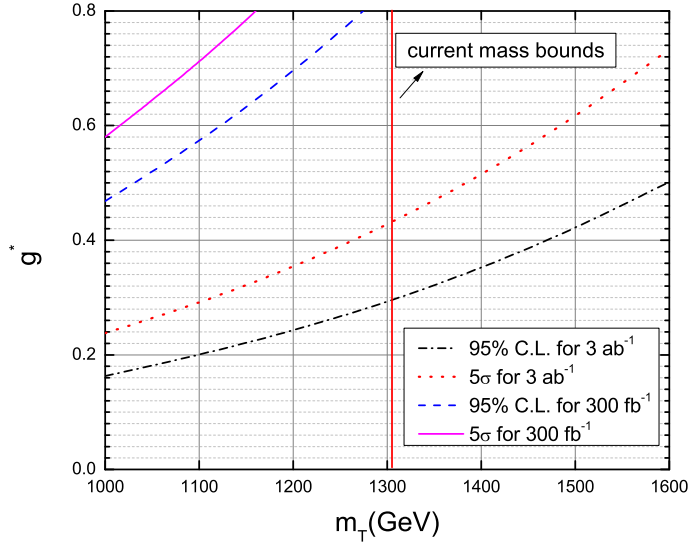}}
\caption{The discovery prospects (at $5\sigma$) and exclusion limit (at $95\%$ CL) for the signal on the ($m_T, g^{\ast})$ plane at the (HL-)LHC with 300 fb$^{-1}$ and 3 ab$^{-1}$, respectively. The red solid vertical line denotes the current bound on the singlet top partner mass at 1310 GeV from Ref.~\cite{13tev-tp2}.}
\label{ss}
\end{center}
\end{figure}

At the HL-LHC, the integrated luminosity is planned to reach 3 ab$^{-1}$, a tenfold increase with respect to the standard LHC.
Using Eqs.~(\ref{eq:ss5})and (\ref{eq:ss2}), we can obtain the expected sensitivity over the place ($m_T,g^{\ast})$ in terms of
 the discovery prospects and exclusion limit of our proposed signals, as shown in Fig.~\ref{ss}, as a function of the top quark partner mass, for these two ${\cal L}_{\rm int}$ values. From this figure we can see that, at the HL-LHC, for $m_T=1.0~(1.2)$ TeV,
the $5\sigma$ level (i.e., $\mathcal{Z}_{E}\geq 5$)  discovery sensitivity on $g^{\ast}$ would be about
 0.24~(0.35), while the upper exclusion limit on $g^{\ast}$  would be 0.16~(0.24) at $95\%$ confidence level (CL) or equivalently with $\mathcal{Z}_{E}\geq 2$. For full luminosity at the standard LHC, $g^*$ values probed are clearly $\sqrt{10}$ higher.
For illustration, the current exclusion limit obtained by the ATLAS Collaboration is 0.29~(0.49) for a singlet $T$ quark of mass of 1.0~(1.2) TeV, using all other available $h$ decay channels.

Note that the latest limits on the singlet top partner mass, assuming a variety of SM-like decay channels [but not $T\to th(\to WW^*)$], imply that all masses below 1310 GeV are  excluded by the
 ATLAS Collaboration~\cite{13tev-tp2}. As shown in Fig.~\ref{ss}, our channel has  no prerogative to enable further sensitivity even at 300 fb$^{-1}$, as the corresponding significance curves in the plot (for perturbative values of $g^*$) lie below the current mass limit. Instead, we are interested here in masses from, say, 1.5 TeV onwards, where the $T\bar T$ channel will be overcome by the single-$T$ one (for certain values of $g^*$), owing to the phase space suppression onto the former, which indeed affects the latter much less (see, e.g., \cite{Liu:2017sdg} and Refs. [30]--[37] therein). Hence, we focus on a top partner with large mass, e.g.,  $m_T=1.5~(1.6)$ TeV,  for which the $5\sigma$ level discovery sensitivity on $g^{\ast}$ would be about 0.62~(0.72), while the upper exclusion limit on $g^{\ast}$  would be 0.42~(0.5) at $95\%$ CL at the HL-LHC.

Certainly, our results can
be applied to other NP models with such top quark partners, such as the minimal CHM of Ref.~\cite{jhep1304-004} with  singlet top quark partners, where the
coset structure is $SO(5)/SO(4)$. The vectorlike top quark partners
can be either in the fourplet or singlet of the unbroken $SO(4)$. In the singlet case, only one $SU(2)$-singlet charged $2/3$ top quark partner is introduced.
From the couplings of the top quark partner with the $W$ boson and a $b$-quark,
the mixing parameter $g^{\ast}$ is given by
\beq
g^{\ast}\simeq \frac{\sqrt{2}y}{g}\frac{m_{W}}{m_{T}},
\eeq
where $y$ is a Yukawa coupling controlling the mixing between the
composite and elementary states. For illustration, with $y=1$ and $m_T=1$~TeV, one obtains $g^{\ast}\simeq 0.17$.

\begin{figure}[!htb]
\begin{center}
\centerline{\epsfxsize=11cm \epsffile{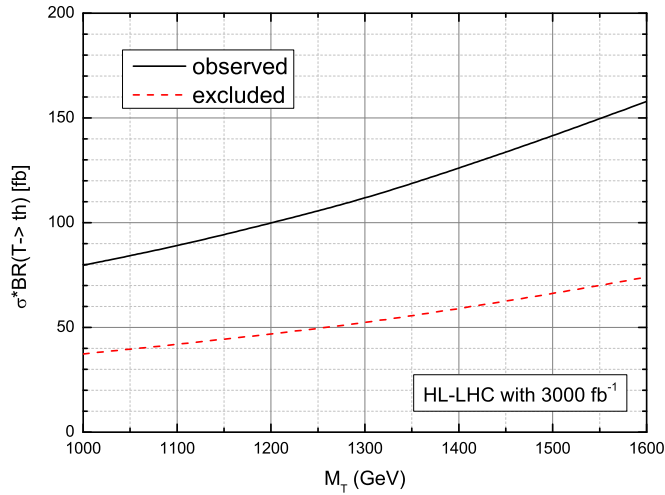}}
\caption{The excluded and observed cross section times BR rates for the single-$T$ signal as a function of the vectorlike top quark partner mass $m_T$ at the HL-LHC. }
\label{CR}
\end{center}
\end{figure}

 Because our results are obtained from fixed BRs in a simplified model, while the latter for different decay channels can be altered in other models, in Fig.~\ref{CR}, we plot the HL-LHC projected sensitivity in terms of the production cross section times BR [$\sigma_T\times {\rm BR}(T\to th)$] as a function of the vectorlike top quark partner mass. We find that  single-$T$  production and decay rates such that $\sigma_T\times {\rm BR}(T\to th)\sim80-160$ fb could be discovered at the HL-LHC for $m_T\in [1.0, 1.6]$ TeV, while the cross sections $\sim37-74$ fb will be excluded.

Before closing, it is interesting to compare our results for the $h\to WW^*$ channel in this specific CHM with some recent ones in the same theoretical context.
In Ref.~\cite{jhep-1604-014}, the authors studied
 search strategies for  single top quark partner production followed by all  possible decay
modes (i.e., $tZ$, $th$ and $Wb$) at the 14 TeV LHC for $m_T=1.0~(1.5)$ TeV. The results show that, with fixed BR$(T\to th)\sim0.25$, the production cross sections of $\sigma_{T+\bar{T}}\sim$130~(55) fb for $m_T=1~(1.5)$ TeV, respectively, could be discovered at the LHC with standard luminosity. Similarly, the cross sections of $\sigma_{T+\bar{T}}\sim$50~(22) fb for $m_T=1~(1.5)$ TeV, respectively, can be excluded. These bounds are therefore more constraining than our results because of the different cut analysis (e.g., they look for a hadronic top decay) and a relatively larger event rate in the $h\to b\bar b$ channel, yet our analysis of the $h\to WW^*$ mode
can represent a complementary candidate to search for a possible singlet top quark partner, at both the standard and HL-LHC.

\section{CONCLUSION}

New heavy vectorlike top quark partners $T$ are predicted in many different NP models, which might then generate a rich phenomenology at the LHC. In this paper, we have studied the prospects of observing  single-$T$ production at the current LHC and future HL-LHC via the $T\to th$ decay channel, followed by a leptonic top  decay and $h\to WW^{*}\to \ell^{+}\ell^{-}+\slashed E_T$.
We performed a model-independent analysis of this process at  $\sqrt s=14 $ TeV  with a simplified model which only includes two free parameters, the top quark partner mass $m_T$ and the EW coupling constant $g^{\ast}$.
The discovery prospects at $5\sigma$ and exclusion limits at $95\%$ CL in the parameter plane of the two variables $m_T$ and $g^{\ast}$ were obtained with both a standard and high luminosity, 300 fb$^{-1}$ and 3 ab$^{-1}$, respectively. For $m_T=1.5~(1.6)$ TeV, our results (at the HL-LHC) show that
the $5\sigma$ level discovery sensitivities of the coupling parameter $g^{\ast}$ are about
 0.62~(0.72), while the exclusion limits at $95\%$ CL on $g^{\ast}$ are given as $g^{\ast}\leq 0.42~(0.5)$.

  Our results can also be mapped over other NP models where the top quark partners only have couplings to the third generation
of SM quarks, e.g., the minimal CHM with singlet top quark partners. In this scenario, we presented the projected sensitivity in terms of the production cross section times BR rates for the $T\to th$ channel. For $m_T\in [1.0, 1.6]$ TeV, rates of $\sigma\times {\rm BR}(T\to th)\sim 80-160$ fb could be discovered while the cross sections $\sim37-74$ fb would be excluded at the HL-LHC.

\begin{acknowledgments}
The work of Y.-B. Liu is supported by the Foundation of Henan Institute of Science and Technology (Grant No. 2016ZD01) and the China Scholarship Council (Grant No. 201708410324). The work of S.M. is supported in part by the NExT Institute and the STFC CG ST/L000296/1.
\end{acknowledgments}


\begin{appendix}
\section{Tree level decay widths}\label{sec:decay}
At the tree level, there are  three top partner decay channels  into SM objects in our scenario, i.e.,
$T\to bW$, $tZ$ and $th$, and the corresponding partial widths are given as
\beq
\Gamma(T\to bW)&=&\frac{\lambda^{\frac{1}{2}}(1, \frac{m^{2}_{b}}{m_{T}^{2}}, \frac{m^{2}_{W}}{m_{T}^{2}})}{32\pi m_{T}}\frac{(gg^{\ast})^{2}}{2}[m_{T}^{2}+m_{b}^{2}-2m_{W}^{2}+\frac{(m_{T}^{2}-m_{b}^{2})^{2}}{m_{W}^{2}}], \\
\Gamma(T\to tZ)&=&\frac{\lambda^{\frac{1}{2}}(1, \frac{m^{2}_{t}}{m_{T}^{2}}, \frac{m^{2}_{Z}}{m_{T}^{2}})}{32\pi m_{T}}\frac{(gg^{\ast})^{2}}{4c_{W}^{2}}[m_{T}^{2}+m_{t}^{2}-2m_{Z}^{2}+\frac{(m_{T}^{2}-m_{t}^{2})^{2}}{m_{Z}^{2}}], \\
\Gamma(T\to th)&=&\frac{\lambda^{\frac{1}{2}}(1, \frac{m^{2}_{t}}{m_{T}^{2}}, \frac{m^{2}_{h}}{m_{T}^{2}})}{32\pi m_{T}}\frac{(gg^{\ast})^{2}}{4m_{W}^{2}}[(m_{T}^{2}+m_{t}^{2})(m_{T}^{2}+m_{t}^{2}-m_{h}^{2})+m_{T}^{2}m_{t}^{2}],
\eeq
where the phase space function $\lambda^{\frac{1}{2}}(a,b,c)=\sqrt{a^{2}+b^{2}+c^{2}-2ab-2ac-2bc}$.

In the large mass limit $m_{T}\gg m_t$, the above expressions can be written as
\beq
\Gamma(T\to bW)&\simeq&\frac{(gg^{\ast})^{2}}{64\pi m_{W}^{2}}m_{T}^{3},\\
\Gamma(T\to tZ)&\simeq&\frac{(gg^{\ast})^{2}}{128\pi m_{Z}^{2}\cos^{2}\theta_{W}}m_{T}^{3},\\
\Gamma(T\to th)&\simeq&\frac{(gg^{\ast})^{2}}{128\pi m_{W}^{2}}m_{T}^{3}.
\eeq
Therefore, in this limit, the BRs of the above three decay modes scale as nearly $2:1:1$.
\end{appendix}

\end{document}